\newif\ifemulateapj
\emulateapjfalse

\ifemulateapj
\documentclass[iop]{emulateapj}

\newcommand{\eqnewline}[1]{\nonumber \\ #1}

\else
\documentclass[12pt,preprint]{aastex}

\newcommand{\eqnewline}[1]{}

\fi

\usepackage{graphicx}
\usepackage{bm}
\usepackage{amsmath}
\usepackage{natbib}
\usepackage{color}
\usepackage{multirow}

\newcommand{\eps}{\epsilon}

\newcommand{\eqb}{\begin{eqnarray}}
\newcommand{\eqe}{\end{eqnarray}}
\newcommand{\diff}{\textrm{d}}

\newcommand{\period}{P_{\rm sec}}

\newcommand{\pperp}{p_{\perp}}

\newcommand{\ppar}{p_{\parallel}}

\newcommand{\gammaw}{\gamma_{\rm w}}

\newcommand{\betaw}{\beta_{\rm w}}

\slugcomment{Submitted to ApJ}
\shorttitle{Emission signatures of waves in pulsar winds}
\begin{document}
\title{Radiative damping and emission signatures of strong superluminal waves in pulsar winds}
\author{Iwona Mochol and John G. Kirk}
\affil{Max-Planck-Institut f\"ur Kernphysik, Postfach 10~39~80,
69029 Heidelberg, Germany}
\email{iwona.mochol@mpi-hd.mpg.de, john.kirk@mpi-hd.mpg.de}

\begin{abstract}

  We analyse the damping by radiation reaction and by Compton drag
  of strong, superluminal electromagnetic waves in the context of
  pulsar winds. The associated radiation signature is found by
  estimating the efficiency and the characteristic radiation
  frequencies.
Applying these estimates to the gamma-ray binary containing PSR~B1259$-$63, we
show that the GeV flare observed by Fermi-LAT can be understood as inverse
Compton emission by particles scattering photons from the companion
star, if the pulsar wind termination shock acquires a precursor of
superluminal waves roughly 30~days after periastron. This constrains
the mass-loading factor of the wind $\mu=L/\dot{N}mc^2$ (where $L$ is
the luminosity and $\dot{N}$ the rate of loss of electrons and
positrons) to be roughly $6\times 10^4$.

\end{abstract}


\section{Introduction}

Pulsar winds contain electromagnetic fields, relativistic electrons,
positrons and possibly ions, and are thought to power the diffuse,
broad band continuum emission observed from pulsar wind nebulae. Their
energetics is dominated by Poynting flux, but the nebulae outside
their termination shocks appear to contain electromagnetic fields and
relativistic leptons approximately in equipartition  \citep[for a recent
review, see][]{2009ASSL..357..421K}.  The implied conversion
process is still not completely understood, although it is becoming
clear that only the wave-like, oscillating components of the fields
must be dissipated at or inside the termination shock, because the
phase-averaged component can dissipate in the bulk of the nebula
\citep{1998ApJ...493..291B,2013MNRAS.431L..48P} without violating observational
constraints.  

Although dissipation of the wave component 
in a freely expanding wind is possible
\citep{2003ApJ...591..366K,2010ApJ...725L.234L}
most current work concentrates on the structure of the shock itself, which is
located close to the point where the ram pressure of the pulsar wind
drops to that of the confining nebula. Two scenarios are under
discussion: In the first, the wave is treated as a series of current
sheets (the \lq\lq striped wind\rq\rq) which are compressed as they
pass through an MHD shock, and subsequently dissipate by driven
magnetic reconnection
\citep{2003MNRAS.345..153L,2007A&A...473..683P,2008ApJ...682.1436L,
  2011ApJ...741...39S,2012CS&D....5a4014S}. In the second, the wave
converts into strong electromagnetic waves of superluminal phase
velocity which dissipate in an extended shock precursor
\citep{2012ApJ...745..108A,2013ApJ...771...53M,2013ApJ...770...18A}.

These scenarios are, in a sense, complementary, since the former
appears to operate at relatively high plasma density, whereas the
latter requires the density to be below a certain threshold. However,
observational constraints on the plasma density in pulsar winds are
loose, and it is not clear which scenario operates in any given
pulsar.  The purpose of this paper is to show that it might be
possible to distinguish between these scenarios observationally,
because they can be expected to have distinct radiation
signatures. The underlying physical reason is that the particles in an
extended electromagnetic shock precursor radiate at relatively low
energy, before thermalisation at a shock front, whereas, in the driven
reconnection scenario, particles are expected to thermalise before
they have a chance to radiate. 

In Section~\ref{emwaves} we summarise the properties of the
electromagnetic waves relevant in a shock
precursor. Section~\ref{radiationdamping} then discusses damping by 
the classical radiation reaction force and by the Compton drag force exerted
when electrons and  positrons moving in the waves scatter ambient photons.
These dissipation processes lead to synchrotron-like emission 
(called synchro-Compton emission) and inverse Compton emission, 
respectively. Estimates of the resulting radiation signature 
are applied in Section~\ref{1259} to the wind from PSR~B1259$-$63,
which is in an eccentric orbit around a luminous Be~star.  This system
is particularly interesting, since the termination shock has the
potential to transit between the driven reconnection and the
electromagnetic precursor scenarios as a function of orbital phase.
We suggest that such a transition can produce a flare similar to the
gamma-ray flare observed by Fermi-LAT
\citep{2011ApJ...736L..11A,2011ApJ...736L..10T} a few days after
periastron.

\section{Electromagnetic precursors and radiation damping}
\label{emwaves}

\subsection{The cut-off radius}

Low amplitude electromagnetic waves can propagate in an
electron-positron pair plasma if their frequency exceeds the local
plasma frequency.  This statement holds
for circularly polarised waves of arbitrary amplitude, provided the plasma 
frequency is defined in
terms of the local phase-averaged {\em proper} density $n$ of the
electron or positron fluid: 
$\omega_{\rm p}=\left(8\pi n e^2/m\right)^{1/2}$. 
In a pulsar wind, waves are driven at the
rotation frequency of the pulsar, and the radial decrease of proper
density leads to the existence of a critical minimum radius 
for electromagnetic wave propagation.

The particle flux density at distance $r$ from the pulsar
and the energy (in units of $mc^2$) carried per particle
in the wind are:
\begin{align}
J&=\dot{N}/\left( r^2 \Omega_{\rm s} \right)
\label{particleflux}\\
\noalign{\hbox{and}}
\mu&=L_{\rm sd}/\dot{N}mc^2,
\label{mudefinition}
\end{align}
where $\dot{N}$ is the e$^\pm$ production rate of the pulsar, 
$\Omega_{\rm s}$ the solid angle occupied by the wind, and
$L_{\rm sd}$ is 
the luminosity carried by the wind, which is
usually assumed to equal the spin-down power.
The parameter $\mu$ corresponds to the maximum possible particle Lorentz factor
$\gamma$, that is achieved only when the entire energy flux is carried by 
cold particles. From the definition of $J$ and $\mu$, and 
the inequalities $J<2\gamma  c n<2\mu c n$, one finds 
\begin{align}
r&> \frac{a_{\rm L} c}{\mu\omega_{\rm p}}
\end{align}
where $a_{\rm L}$ is a --- typically very large --- 
dimensionless parameter related to the luminosity
per unit solid angle:
\begin{align}
a_{\rm L}&=\sqrt{\frac{4\pi e^2 L_{\rm sd}}{m^2c^5\Omega_{\rm s}}}
\label{aldefinition}\\
&=3.4\times 10^{10} \left(L_{\rm sd}/10^{38}\textrm{erg\,s}^{-1}\right)^{1/2}
\left(\Omega_{\rm s}/4\pi\right)^{-1/2}
\nonumber
\end{align}
This suggests the use of a dimensionless radius coordinate:
\begin{align}
R&=r\omega\mu/c a_{\rm L}
\label{normalize}
\end{align}
and a detailed treatment indeed reveals that the cut-off radius
for both linear and circular polarised waves
lies very close to, but slightly outside 
$R=1$ \citep{2012ApJ...745..108A}.

\subsection{Subluminal modes}

For $R<1$, it is expected that the MHD approximation is adequate to
describe the plasma dynamics. The oscillating components of the fields
can be assumed to be carried along with the plasma in a
quasi-stationary pattern containing a combination of current sheets
and/or magnetic shear, all moving at subluminal phase speed.  But this
subluminal mode is not restricted to small radius; it exists out to
arbitrary large distance.  Its properties depend on $\mu$ and on the
magnetisation parameter $\sigma_0$, defined as the ratio of Poynting
flux to kinetic energy flux. A pulsar wind is expected to be cold,
supermagnetosonic and Poynting-flux dominated when launched, which
implies $\mu\gg1$ and $\sigma_0\lesssim\mu^{2/3}$. The radial momentum
per particle, $\nu mc$, is in this case almost equal to the 
energy carried radially per particle, divided by $c$:
\begin{align}
\nu&\approx\mu-\frac{\sigma_0}{2\mu}.
\end{align}
Subluminal modes emit very little
radiation, since the particle trajectories in them are almost ballistic.  An
observable signal can be expected only from inverse Compton scattering in the presence of a dense
ambient photon field \citep{2000APh....12..335B,2000MNRAS.313..504B,2008A&A...488...37C}. 
The magnetisation
stays close to its initial value up to very large
distance $R\sim \mu/\sigma_0$, after which it gradually falls off, 
as the Poynting flux converts to
particle energy. This phase may be relevant in a charge-starved 
blazar jet \citep{2011ApJ...736..165K} but it is not 
likely to
be encountered in a pulsar wind, which is confined by either the
interstellar medium or the wind of a companion star.

\subsection{Jump conditions and superluminal modes}
At sufficiently large radius, the MHD
approximation, which excludes electromagnetic waves of superluminal phase 
speed, can fail.  In particular, if the termination shock of a pulsar wind
lies at $R>1$, a precursor dominated by these waves can
form. The dynamics of such a structure is complex, containing both
forward and backward propagating waves
\citep{2013ApJ...770...18A}, and it is not clear that a steady state is
established. However, in order to gain insight into the possible
configurations, and the radiation signature they might emit, it is
useful to study a highly simplified case that is accessible
analytically. In this paper, therefore, we concentrate on a
solution in which the precursor is described by a steady, circularly 
polarised wave in a cold, two-fluid (electron and positron) model.

Initially, this wave must carry the 
same particle, energy and radial momentum fluxes as the incoming
subluminal mode. This leads to a set of jump conditions
that determine the phase-averaged wave quantities 
at launch \citep{2012ApJ...745..108A}:
\begin{align}
J&=2c n\ppar 
\label{jumpconditions1}\\
\mu&=\left(\gamma+\frac{\betaw\gammaw^2\pperp^2}{\ppar}\right)
\label{jumpconditions2}\\
\nu&=\left(\ppar +\frac{\left(1+\betaw^2\right)\gammaw^2\pperp^2}{2\ppar}
\right)
\label{jumpconditions3}
\end{align}
where 
$\gamma$, $\ppar$ and $\pperp$ are the Lorentz factor, the 
parallel momentum and the magnitude of the transverse momenta of the 
fluids 
(the latter both in units of $mc$). Together with the proper density $n$,  
these quantities are phase-independent. The wave group speed 
 $\betaw$ ($=ck/\omega$) and corresponding Lorentz factor,
$\gammaw$,
follow from the dispersion relation 
\begin{align}
\omega^2&=\omega_{\rm p}^2+c^2k^2.
\end{align}
The amplitude $E$ of the wave electric field, which is purely transverse, 
is related to the transverse momentum
\begin{align}
E&=mc\pperp\omega/e
\end{align}
and can take on arbitrarily large values, as can also $\ppar$. Note, however,
that for a given lab.\ density, these quantities enter into the definition of 
the plasma frequency, allowing large amplitude waves to propagate through
dense plasmas. 

For a given set of parameters $\mu$ and $\sigma_0$ (and, hence $\nu$) of
the incoming subluminal mode there are, at each radius outside the 
cut-off, two
solutions to the jump conditions
(\ref{jumpconditions1}--\ref{jumpconditions3}), corresponding to a
free-escape and a confined mode.  In a shock precursor, the latter is
the relevant solution.  Except very close to $R=1$, it is
characterised by a relatively large transverse momentum. This can be
exploited to find an approximate solution. From the continuity equation
(\ref{jumpconditions1}), and the definitions (\ref{particleflux}), 
(\ref{mudefinition}) and (\ref{aldefinition}), it follows that:
\begin{align}
\ppar&=\gammaw^2\mu/R^2.
\end{align}
Substituting this expression into (\ref{jumpconditions2}) and 
(\ref{jumpconditions3}) and observing that, for $\ppar\ll\gamma$,
the particle contribution to 
the momentum flux is negligibly small, one finds, provided the flow
remains relativistic ($\gamma\gg1$),
\begin{align}
\gamma&\approx\pperp
\label{approxjump1}\\
&\approx \frac{\mu\left(1-\betaw\right)^2}{1+\betaw^2}
\label{approxjump2}\\
\ppar&\approx\frac{\mu\left(1-\betaw\right)^3}{2\left(1+\betaw\right)
\left(1+\betaw^2\right)}
\label{approxjump3}\\
R&\approx\frac{2^{1/2}\left(1+\betaw^2\right)^{1/2}}{\left(1-\betaw\right)^2}.
\label{approxjump4}
\end{align}
Thus, all quantities at the launching point are approximately determined by the single parameter $R$, or, 
alternatively, $\betaw$.

\section{Radiation damping}
\label{radiationdamping}
Since the precursor extends over a distance that is at most 
comparable to the 
radius of the termination shock, 
it can be regarded for the purposes of our estimates as 
a plane wave. 
The damping effect of radiation can then 
be treated using 
a perturbative approach, as 
originally presented for the linearly polarised case by 
\citet{1978A&A....65..401A}. 
The procedure is closely related 
to that described for 
spherical wave propagation in the short wavelength approximation 
described in \citet{2013ApJ...771...53M}. 
In the ultra-relativistic limit, the radiation-reaction force is almost 
anti-parallel to the particle momentum, and has time-like and $x$-components
\citep{1975ctf..book.....L}
\begin{align}
g^{0,1}&\approx-\frac{2e^4}{3m^3c^6}E^2\left|\gamma -\betaw\ppar\right|^2(\gamma,\ppar)\enspace.
\end{align}
Equations~(9) and (10) of \citet{2013ApJ...771...53M}, simplified to
the case of circular polarisation (in which case the phase-averaging
is trivial) and to planar geometry, acquire the terms $ng^0$ and
$ng^1$ on their respective right-hand sides. The energy and entropy
equations --- (9) and (12) of that paper --- then take the form
\begin{align}
\frac{d}{dx}\left(2n\ppar\gamma+\frac{\betaw E^2}{4\pi mc^2}\right)&=ng^0  \\
\frac{d\gamma}{dx}&=\frac{g^1-\betaw g^0}{2\Delta} 
\end{align}
By introducing 
the space-dependent magnetisation parameter
\begin{align}
\sigma&=\frac{\betaw\pperp^2 R^2}{\mu\gamma}\\
\noalign{\hbox{and using the definitions}}
\delta&=\ppar-\betaw\gamma\\
\Delta&=\gamma-\betaw\ppar\\
\eps&=\frac{2 e^2\omega}{3mc^3}.
\end{align}
these can be written as
\footnote{%
We note that the entropy equation given 
by \citet{1978A&A....65..401A} (eq~32 in that paper) --- 
contains an error arising from the neglect
of first-order terms in Faraday's equation (eq~14).}
\begin{align}
\frac{2\mu}{\eps a_{\rm L}}\frac{\diff}{\diff X}
\left[\gamma\left(1+\sigma\right)\right]
&=-\frac{\gamma\Delta^2\pperp^2}{\ppar}
\label{energyloss}\\
\frac{2\mu}{\eps a_{\rm L}}\frac{\diff\gamma}{\diff X}
&=-\delta\Delta\pperp^2
\label{entropyeq}
\end{align}
where the cartesian coordinate along the propagation direction $x$
has been normalised to the cut-off radius, in analogy with 
Eq.~(\ref{normalize}): $X=x\omega \mu /\left(c a_{\rm L}\right)$.

Note that $\sigma$, 
$\delta$, $\Delta$, $\gamma$, $\ppar$, $\pperp$ and $\betaw$ are all
functions of $X$, although the argument has been omitted to
simplify the notation. 
For superluminal waves, $\Delta>0$, implying that the 
Poynting flux, which is proportional to $\gamma\sigma$,
decreases monotonically with $X$ for positive particle flux ($\ppar>0$):
\begin{align}
\frac{2\mu}{\eps a_{\rm L}}\frac{\diff}{\diff X}
\left(\gamma\sigma\right)
&=-\frac{\Delta\pperp^2\left(1+\pperp^2\right)}{\ppar}
\label{poyntingloss}
\end{align}
On the other hand, radiative reaction causes the particle Lorentz
factor $\gamma$ to increase when the wave moves faster than the
particles, i.e., when $\ppar/\gamma-\betaw=\delta/\gamma<0$, but
decrease when the particles are faster.

Equations (\ref{poyntingloss}) and (\ref{entropyeq}) show that the
ratio of the rate of change of Poynting flux to the rate of change of
$\gamma$ is $(1+\pperp^2)/(\ppar\delta)$, which is large except very
close to the point $R=1$.  Thus, on launch, the wave starts to convert
Poynting flux into radiation, whilst leaving the Lorentz factor of the
charged fluids, and, therefore, the flux of kinetic energy unchanged.
This reduction in Poynting flux is attributable entirely to a decrease
of the wave group speed $\betaw$, or, equivalently, an increase of the
phase speed and of the wavelength of this mode.  Ultimately, when the
wavelength approaches the radial distance from the pulsar, the short
wavelength (plane-wave) approximation used to derive the governing
equations breaks down.  
At this point, however, the incoming Poynting
flux has, to a good approximation, been completely converted into
radiation.  The (dimensionless) length-scale $X_{\rm diss}$ on which
this happens follows from eq (\ref{poyntingloss}):
\begin{align}
X_{\rm diss}
&=
\frac{2\mu}{\eps a_{\rm L}}\left(\frac{\gamma\sigma\ppar}{\Delta\pperp^2\left(1+\pperp^2\right)}\right)
\end{align}
where the values of $\gamma$, $\sigma$, $\ppar$, $\pperp$ and $\Delta$
are those at launch, which follow from the jump conditions.  When the
dissipation length exceeds the radial distance to the pulsar, the
plane-wave approximation breaks down well before the Poynting flux has
been converted into radiation. To take account of this, we estimate
the efficiency $\eta$ of conversion of spin-down power into radiation
in the precursor wave as follows:
\begin{align}
\eta&=
\left\lbrace
\begin{array}{ll}
\sigma/(1+\sigma)  & \textrm{for\ }X_{\rm diss}<R\\
\sigma R/\left[X_{\rm diss}(1+\sigma)\right]  & \textrm{for\ }X_{\rm diss}>R
\end{array}
\right.
\label{efficiencydef}
\end{align}

In addition to damping by the radiation reaction term in the fluid
equation of motion, damping of waves in a pulsar wind can also result
from Compton drag --- the force that results from the scattering of
individual, ambient photons by the relativistically moving electrons
and positrons that make up the two cold fluids. In general, this
effect can be quite complicated, since it depends on both the angular
and spectral distributions of the target photon field.  Ambient photon
fields such as the light from distant stars, the cosmic microwave
background or synchrotron radiation from the pulsar wind nebula can be
assumed to be isotropic, which considerably simplifies the
computations. However, photons radiated by the neutron star surface or
originating from a close binary companion are highly anisotropic,
which has important consequences for radial winds
\citep{1999APh....10...31K,2000MNRAS.313..504B}.  
Nevertheless, since
our treatment is restricted to an electron velocity with a large
angular spread, $\pperp/\ppar\gg1$, the average drag is approximately
equal to that found assuming the target photons to be isotropic.

In pulsar winds, the target photons, whose characteristic 
frequency will be denoted by $\nu_0\equiv x_0mc^2/h$,  
can be assumed to be soft, and 
the drag force is approximately anti-parallel to the fluid velocity, 
as it is in the case of radiation reaction. For scatterings that
take place in the Thomson regime, $\gamma x_0<1$, the 
computation is relatively simple. 
However, 
in the case of 
gamma-ray binaries, the drag exerted by photons
from the companion star is reduced significantly by the Klein-Nishina effect.
The resulting expression for the dissipation length 
$X_{\rm diss}$ including both the radiation reaction term and the Compton drag is
\begin{align}
X_{\rm diss}&=\frac{2\mu}{\eps a_{\rm L}}\left(\frac{\gamma\sigma\ppar}{\Delta\pperp^2\left(1+\pperp^2\right)}\right)
\left(1+\frac{G(\gamma,x_0)\gamma_{\rm ic}^2}{\Delta^2}\right)^{-1}
\label{xdissdef}
\end{align}
where $G(\gamma,x_0)$ is the reduction factor due to Klein-Nishina
effects ($G(\gamma,x_0)\rightarrow1$ for $\gamma x_0\rightarrow0$ such
that $\gamma x_0^{1/3}\rightarrow\infty$) originally given by
\citet{1965PhRv..137.1306J}, see also
\citet[][equation~(A10)]{1999APh....10...31K}, and the energy density
of target radiation $U_{\rm rad}$ enters via an effective Lorentz
factor $\gamma_{\rm ic}$ at which the two terms have approximately
equal magnitudes
\begin{align}
  \gamma_{\rm ic}&=\left(\frac{2\sigma_{\rm T} U_{\rm rad}
      c^2}{e^2\omega^2}\right)^{1/2}
\end{align}
with $\sigma_{\rm T}$ the Thomson cross section.

Adopting the energy density of the cosmic microwave background,
$\gamma_{\rm ic}=7.3\times P_{\rm sec}$, where $P_{\rm sec}$ is the
pulsar period in seconds. Thus, provided the termination shock is
located sufficiently close to the pulsar to produce relativistic
particles in the precursor ($R\ll \mu$), inverse Compton scattering
with this radiation field as a target is energetically unimportant.
In this case, the condition for a substantial fraction of the
spin-down power to be converted into radiation, $X_{\rm diss}<R$,
reads:
\begin{align}
\frac{1}{2}\left(\eps a_{\rm L}\mu^2\right)f(R)
&>1
\label{dissipscale}
\end{align} 
where
\begin{align}
f(R)&=\frac{\Delta\left(1+\pperp^2\right)}{\betaw\ppar\mu^2 R}
\label{functionf}
\end{align} 

\begin{figure}
\input{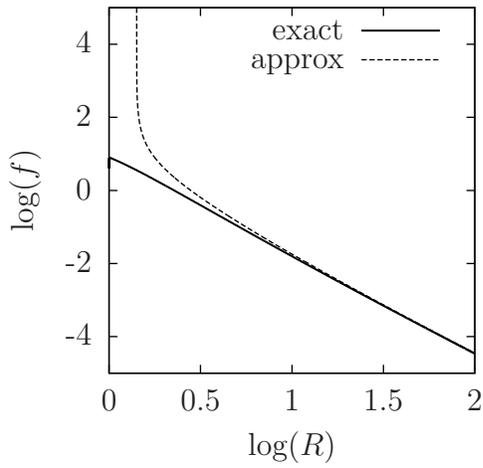}
\caption{\label{figurefg}
The function $f(R)$ defined in Eq.~(\ref{functionf}) 
that determines the efficiency of synchro-Compton radiation, 
as a function of the dimensionless radius $R$. 
The curves labelled \lq\lq exact\rq\rq\ 
and \lq\lq approx\rq\rq\ are calculated using the exact and approximate jump 
conditions, defined by Eqs~(\ref{jumpconditions1}--\ref{jumpconditions3})
and Eqs~(\ref{approxjump1}--\ref{approxjump4}) respectively.}
\end{figure} 

The function $f$ is plotted in Fig.~\ref{figurefg}. 
For relativistic flows, it is a function of $R$ only, as indicated. Using the 
approximate jump conditions (\ref{approxjump1}--\ref{approxjump2}), according
to which $\gamma\rightarrow\mu/R$
and $\ppar\rightarrow\mu/\left(2R\right)^{3/2}$ for large $R$,
one finds
\begin{align}
f(R)&\approx 2\sqrt{2}R^{-5/2}\ \textrm{for\ }R\gg1
\end{align}
which demonstrates that the importance of radiation losses falls off
rapidly as the termination shock radius becomes large.  However, close
to the cut-off radius, $f\sim1$, so that, assuming the termination
shock lies close to this point, i.e. at
\begin{align}
r&\sim 10^{16} L_{38}^{1/2}P_{\rm sec}\mu_4^{-1}\,\textrm{cm}
\end{align}
(where $\mu=\mu_4\times10^4$), 
the condition for 
the precursor to radiate a significant fraction of the wind luminosity 
as a result of damping by radiation reaction is
roughly 
\begin{align}
\eps a_{\rm L}\mu^2&=
1.3\times10^{-4}\mu_4^2 L_{38}^{1/2}\period^{-1}\\
&>1.
\label{ineq}
\end{align}

With Compton drag included in the definition of $X_{\rm diss}$, the
radiation efficiency (\ref{efficiencydef}) can be split into two
terms. The first, $\eta_{\rm sc}$, arises from radiation reaction and
channels energy into synchrotron-like emission produced by the
curvature in the fluid trajectories, a process that has been called
\lq\lq synchro-Compton\rq\rq\ emission \citep{1971IAUS...46..407R}.
The second term, $\eta_{\rm ic}$, arises from Compton drag and
channels energy into inverse Compton scattered photons:
\begin{align}
\eta_{\rm sc}&=\eta/\left(1+\frac{G(\gamma,x_0)\gamma_{\rm ic}^2}{\Delta^2}\right)
\label{etascdef}
\\
\eta_{\rm ic}&=\eta/\left(1+\frac{\Delta^2}{G(\gamma,x_0)\gamma_{\rm ic}^2}\right)
\label{etaicdef}
\end{align}

In the idealised, cold, two-fluid model used here, both radiation
processes produce photons in narrow spectral bands.  The
synchro-Compton emission peaks at a (dimensionless) photon energy 
given by
\citep[e.g.,][]{1972A&A....20..135B}
\begin{align}
\bar{x}_{\rm sc}(\gamma,\pperp)&=0.4\gamma^2\pperp \left(\hbar\omega/mc^2\right),
\label{xscdef}
\end{align} 
whereas the peak energy of inverse Compton scattered photons can
be estimated as
\begin{align}
\bar{x}_{\rm ic}(\gamma,x_0)&=(4/3)\sigma_{\rm T}x_0 \gamma^2 G(\gamma,x_0)
/\langle \dot{N}_\gamma\rangle,
\label{xicdef}
\end{align}
where $\dot{N}_\gamma$ is the scattering rate divided by the density of
target photons and $\langle\dots\rangle$ indicates an angle average,
i.e.,
\begin{align}
\langle\dot{N}_\gamma\rangle&= \frac{c}{2\gamma x_0}
\int_{\gamma x_0(1-\beta)}^{\gamma x_0(1+\beta)}\,\diff x'\,x'\sigma_{\rm KN}\left(x'\right)
\end{align}
where $\beta=\sqrt{\gamma^2-1}/\gamma$, and $\sigma_{\rm KN}(x)$ 
is the total Klein-Nishina cross section for a
photon of energy $xmc^2$
\citep[see][]{1965PhRv..137.1306J,1999APh....10...31K}.

\section{Application to PSR~B1259$-$63/SS2883}
\label{1259}

The radiation efficiencies, determined by Eqs~(\ref{efficiencydef}), 
(\ref{xdissdef}), (\ref{etascdef}) and (\ref{etaicdef}), together with 
the peak photon energies (\ref{xscdef}) and (\ref{xicdef}) may be used to 
estimate the radiation signature of a potential electromagnetic
precursor to the  pulsar wind termination shock, 
given the pulsar's angular frequency
$\omega$, the parameter $a_{\rm L}$, defined in eq~(\ref{aldefinition}), 
the distance $r_{\rm ts}$ of the shock from the pulsar, the energy density and
characteristic frequency of the dominant soft photon targets 
and the mass-loading parameter $\mu$. 

In this section we use pulsar PSR~B1259$-$63 as an illustrative
example. This intensively observed and modelled object is a member of
the gamma-ray binary class \citep{2006A&A...456..801D}, and was the
first source to be discovered in TeV gamma-rays by the HESS
collaboration \citep{2005A&A...442....1A}. Its angular frequency is
$131\,\textrm{s}^{-1}$ and its spin-down luminosity is
$8.3\times10^{35}\,\textrm{erg\,s}^{-1}$, giving $a_{\rm
  L}=3\times 10^{9}$, assuming the pulsar wind occupies a solid angle of
$4\pi\,$sr.  The 3.4~yr orbit is eccentric, the separation between the two
stars varying by roughly a factor of 14 between apastron and
periastron, where it is approximately $10^{13}\,\textrm{cm}$.  Thus,
the distance from the pulsar to that part of the termination shock
that lies between the two stars is fairly tightly constrained. The
dominant photon field that serves as a target for inverse-Compton
scattering is provided by the luminous Be~star SS2883. This
leads to $x_0\approx10^{-5}$ and
\begin{align}
\gamma_{\rm ic}&=1.4\times10^6 \left(\frac{L_*}{2.3\times10^{38}\,\textrm{erg\,s}}\right)^{1/2}
\left(\frac{d}{10^{13}\,\textrm{cm}}\right)^{-1}
\label{urad}
\end{align}
where $d$ is the distance to the star and $L_*$ its luminosity, and we  have
adopted as fiducial parameters those given by
\citet{2011ApJ...732L..11N}, which indicate a somewhat more luminous and distant
star than previously thought \citep[e.g.,][]{1994MNRAS.268..430J}.

The most difficult parameter to 
constrain observationally is the mass-loading $\mu$. 
It is related to the \lq\lq multiplicity\rq\rq\ 
$\kappa$ associated with the pair creation mechanism close to the pulsar by
$\mu=a_{\rm L}/4\kappa$, but this parameter is also difficult to estimate. 
Modelling of the nonthermal emission near periastron suggests 
$\mu\approx10^6$ \citep{1999APh....10...31K,2012ApJ...753..127K}, and
identifying the spectral break in keV range with the low energy
cut-off of the synchrotron emission associated with the relativistic
wind suggests a similar value $\mu\approx 4\times10^5$
\citep{2009ApJ...698..911U}.  These rough estimates should
be considered as upper limits, since they do not take into account the 
possibility that 
some particles in the wind might not be accelerated to high energy 
at the shock.
Taking them at face value, places the distance
from the pulsar of the cut-off point for electromagnetic waves ($R=1$)
at $2.3\times10^{12}\,\textrm{cm}$ and $6\times10^{12}\,\textrm{cm}$,
respectively, close to the stand-off distance of the shock from the
pulsar at periastron. A low energy particle component would imply
a larger cut-off radius, so that, close to periastron,
the pulsar wind termination shock might not have an electromagnetic
precursor, but could form one when, as the stars move apart,
the confining pressure of the companion's wind drops, and the shock
emerges through the cut-off radius. If this happens, one should expect a
new emission component to emerge, with the properties described in the
previous section.

Around periastron passage of~2010, the system was detected by
Fermi-LAT \citep{2011ApJ...736L..11A,2011ApJ...736L..10T}.  The most
interesting aspect of these observations is a flare of 
$1\,$GeV photons detected about 30 days after the periastron passage. 
This emission forms a new, spectrally distinct component that carries
a substantial fraction of the pulsar's spin-down luminosity. It presents a 
challenge to models based on electrons accelerated at the termination shock
\citep{2012ApJ...753..127K}, which assume strong Doppler-beaming, as 
well as to those based on electrons in the unshocked wind
\citep{2011MNRAS.417..532P,2011ApJ...742...98K,2012ApJ...752L..17K}, which 
require an additional source of target photons.  In an electromagnetic
precursor, on the other hand, emission is beamed into the plane transverse to
the flow and covers a large fraction of the sky. Also, the
relatively low energy ($\gamma\sim10^4$) electrons and positrons
remain in the emission region for a time that is a factor
$\gamma/\ppar\approx \left(8R\right)^{1/2}$ longer than that of
radially propagating particles, thus enhancing the radiated power.

\begin{figure}
\input{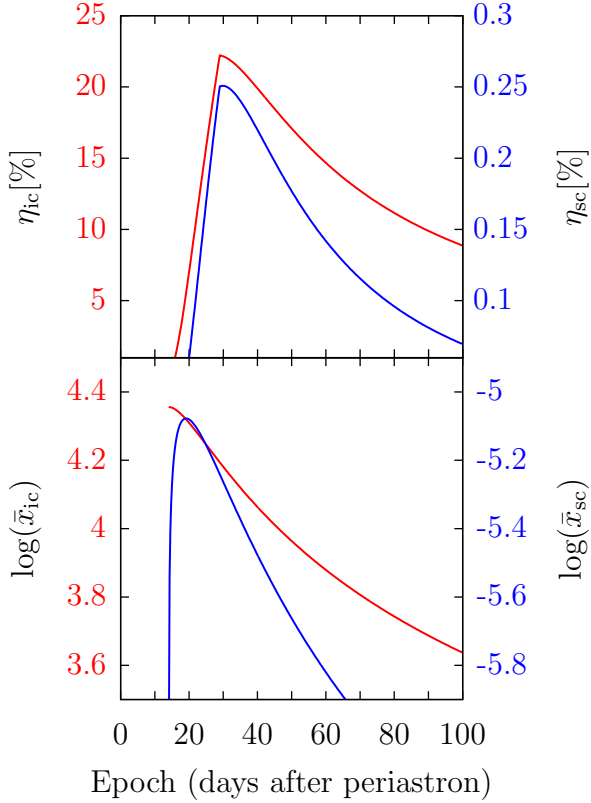}
\caption{\label{losses}The efficiencies and peak energy bands of 
synchro-Compton (blue curves) and inverse Compton
emission (red curves) 
from an electromagnetic shock precursor, calculated for
$\mu=6\times 10^4$ and $\sigma=100$, as a function of the time (in days)
after periastron for the binary system containing PSR~B1259$-$63.
The stellar and orbital parameters are taken from 
\citet{2011ApJ...732L..11N} and the shock is assumed to be located 
mid-way between the two stars.}
\end{figure}

In Fig.~\ref{losses} we plot the efficiencies $\eta_{\rm sc}$ and
$\eta_{\rm ic}$ together with the peak photon energies $\bar{x}_{\rm
sc}$ and $\bar{x}_{\rm ic}$ as functions of time after periastron,
assuming the shock to be located mid-way between the two stars.
The rapid switch-on apparent in this figure is 
somewhat arbitrary, since (i) we assume the precursor is 
created instantaneously with a linear extent comparable to the shock radius
and (ii) the exact jump conditions were used in the calculation, for which an 
estimate of the wind
magnetisation parameter is needed --- we adopt $\sigma_0=100$. 
However, these uncertainties apply only during the rising phase of emission, 
when the shock is close to the cut-off radius.
After the peak of emission at day~30, when the dissipation length exceeds the 
precursor size, the approximate jump conditions, which do not depend 
on $\sigma_0$ are an accurate approximation. Thus, the only unconstrained parameter, $\mu$, is fixed by the 
epoch of maximum radiative efficiency.  To match this to the 
peak of the flare detected by Fermi-LAT, we choose $\mu=6\times 10^4$. 

Between 30 and 100 days after periastron, Fig.~\ref{losses} predicts a
peak energy for the inverse-Compton photons that drops from 
a few GeV to about $1\,$GeV, and a corresponding power that drops from
roughly 20\% to 10\% of the spin-down luminosity.  The synchro-Compton
emission lies in the optical band, but accounts for only a small
fraction of the spin-down luminosity, and would be swamped by photons
from the companion star. Given the highly simplified treatment used to
make these estimates, such as the assumption of a spherically
symmetric wind and of uniform conditions with vanishing phase-averaged
fields over the entire shock front, these properties are in
encouraging agreement with those of the observed flare. 

If the electromagnetic precursor is responsible for the Fermi-LAT
flare, then the emission should be present at all binary phases during
which the termination shock lies outside the cut-off radius. Over most
of the binary orbit, the efficiency would be too low to permit
detection.  In Fig.~\ref{losses} it is assumed that the shock lies
mid-way between the two stars. If this holds over the entire orbit,
and if the mass-loading parameter remains constant, then an additional
flare should occur between 100 and 30 days {\em before} periastron
passage.  This is ruled out by Fermi observations.  However, the
location of the shock front depends on the relative strengths of the
pulsar and the companion star winds, which can be expected to differ
in the pre- and post-periastron parts of the orbit, since the wind of
the companion star is known to be highly anisotropic. Consequently,
although a pre-periastron flare is predicted, and should be described
by the same formalism, it is not necessarily expected to be
symmetrically timed with respect to periastron, and, therefore, not
necessarily peaked in the GeV band.

\section{Conclusions}

Radiation reaction and Compton drag can be important damping
mechanisms of superluminal waves in the winds of short period pulsars,
if they exist close to the cut-off radius. This damping
extracts the Poynting flux from the wind 
directly by reducing the group velocity of
the wave, using the particles at catalysts, i.e., without depleting
the kinetic energy flux. We have derived explicit expressions for the
efficiency of extraction and for the spectral bands into which the energy is
channelled, as functions of the pulsar wind parameters and the properties
of the ambient photon field.

In the case of the gamma-ray binary containing the pulsar
PSR~B1259$-$63, all of these parameters except the mass-loading $\mu$
are well-constrained. Assuming the waves are generated as precursors
to the termination shock of this pulsar when it encounters the wind of
the companion star, the mass-loading parameter determines the epoch of
maximum extraction efficiency. Choosing $\mu=6\times10^4$ fixes this
to coincide with the enigmatic gamma-ray flare detected from this
object by Fermi-LAT. In this case, the overall efficiency and the peak
frequency of the inverse Compton emission are in rough agreement with
those observed, suggesting that the flare is caused by the emergence
of the shock from the cut-off radius, accompanied by the creation of a
precursor containing superluminal waves. This interpretation
predicts another flare when the precursor disappears, but its timing
depends on unknown properties of the wind of the companion star.

\end{document}